\documentstyle[psfig]{lamuphys}
\makeatletter
\let\chapter\hid@chapter
\makeatother
\begin{document}
\pagenumbering{arabic}
\title{Weak lensing at the limit of the sky background noise}

\author{Yannick\,Mellier\inst{1,2} and Bernard\,Fort\inst{2}}

\institute{Institut d'Astrophysique de Paris CNRS, 98 bis Boulevard
Arago,
75014 Paris, France 
\and
Observatoire de Paris,
DEMIRM,
61 Avenue de l'Observatoire, 75014 Paris, France}

\maketitle

\begin{abstract}

Recent weak lensing observations have pushed the  use of 4 meter-class 
telescopes to the limits of their capabilities with exposure times
exceeding  several  hours.  The leading idea is that the surface  
density of faint galaxies up to very faint magnitude ($B > 28-30$)
raises continuously thus potentially offering a dense template of distant 
sources whose intensity contrast is at the sky noise level.  In complement 
to the Peter Schneider's presentation  on dark matter search from weak lensing 
(this conference), we  review some of these recent advances in weak lensing 
analysis based on this extreme faint population of galaxies in order to explore:
\begin{enumerate}
\item the dark matter distribution on large scales,
\item the redshift ditribution of lensed sources at very large distance,
\item and eventually the values of cosmological parameters.
\end{enumerate}
For each observational topic we will briefly 
discuss these new methods as compare to
more classical lensing studies as well as the possible VLT 
scientific impact in the domain. 
\end{abstract}
\section{Cosmology with gravitational lenses}

Gravitational lenses effects on distant galaxies can potentially
probe either the dark matter distribution from low mass compact objects to
large-scale structures, or the angular distances of sources that
depend on their redshift distribution  and to a smaller extend on   
the cosmological parameters.  This has motivated a lot of observational 
efforts devoted to large multiple arcs and arclets in rich clusters of 
galaxies (see Fort \& Mellier 1994 and Narayan \& Bartlemann 1996 for reviews).
Indeed, we still expect important 
results from the strong lensing regime 
with the HST and the VLTs  since recent improvements in 
instrumentation or data analysis provide unprecedented
observational capabilities with 
these facilities. But except in the core of  
compact clusters of galaxies, arc(let)s are relatively rare events 
which cannot neither easily represent a large and fair sample of faint distant 
galaxies at large redshifts (luminosity bias) nor probe the mass distribution
outside high condensation of masses. In this domain the greatest
hope could be that HST observe an exceptional
event with  several multiple images systems at different redshifts that will 
allow a unique test on the geometry of the Universe.

So, after the pioneer analysis of Tyson et al (1990) on the shear field 
around A1689 and the extensive work of theoreticians (see Schneider,
this conference), it  appears clear that weak lensing can  in principle be 
observed everywhere in the sky if we have very deep CCD images of the galaxy 
population at large redshift: the densest the population of background
galaxies, the highest the visibility of the shear effect (better 
angular resolution and/or higher signal to noise ratio). Simultaneously 
crucial questions raised 
to observers: how deep is it possible to go in order to detect a large number 
of sources that can be statistically used to measure a coherent lensing 
signal on a given sky aperture with the highest signal-to-noise ratio? 
Is it actually possible to correct all the 
atmospheric and instrumental distortions of astronomical images whose 
amplitudes can be 5 to 10 times larger than the predicted amplitude 
of the gravitational lensing signal?

A technical answer to the second question is not trivial and far beyond the 
scope of this paper because all the ground based telescopes are affected by 
seeing effect and have been so far constructed without such an ultimate image 
quality in mind. Without a strong and dedicated effort on 
image correction of non axi-symmetric geometrical Point Spread Function, 
the measurement of 
weak shear below a few percent may be only possible with the HST on a few 
limited field of view for a while. This important remark has to be kept in mind
all along the reading of this paper because we implicitly suppose that
the problem has been solved.

For the first question, we know that  for long exposure time with 
excellent seeing\footnote{The 
meaning of this is not clearly defined. Let say that it corresponds
basically to 3-5 hours on a 4 meter telescope with a median seeing 
below 0.7 arcsec.}, the photometry of the faintest galaxies is limited to
about $B=26.5$ 
or $I=25$. Beyond these magnitudes, the sources are still there but hidden 
in the photon noise of the sky background. The good linearity of CCD gives in 
fact a unique opportunity to detect this underground population
and to measure their number density or some  global
coherent geometrical feature, like the weak shear induced by the deviation of
light by condensation of mass. 
In the following we show first how the pixel to pixel autocorrelation
function of the sky background reveals the existence of such 
a large population of 
extremely faint galaxies and how it is possible to use them  to map 
the shear with an unprecedented signal to noise ratio. Then we discuss  
the possibility claimed by Fort et al (1996a)  to detect
the possible positions  of  
sources down to 0.5$\sigma$  of the 
sky level  ($B=28$) and to use the number counts
to study the magnification bias effect around cluster of 
galaxies. Such an approach open a new way   
to explore the redshift distribution of galaxies to limiting
magnitudes far beyond the other methods (spectroscopy and lensing 
inversion), and to search for  new constraints on the value of the 
cosmological constant. 

In this  review, we focus on these recent developements which 
use faint sources at the noise level. Though still in their infancy,    
they will demonstrate their full scientific interest 
when used with the outstanding capabilities of the VLTs.

\section{The autocorrelation function of pixels}

All standard  methods which are used to determine 
the projected mass density from 
a lensing inversion of the shear map proceed in the same basic way.
Very faint objects are detected down to a threshold limit 
that strongly depends on the seeing. Then, the center, shape, 
size and magnitude of every sources are calculated and averaged
within a given solid angle (scanning aperture
on the sky) that defines the angular resolution of 
the shear map. The averaged ellipicity is finally linked to
the potential and projected mass distribution on the sky.
Though pratical 
and easy to implement, the method depends on the detection threshold, 
 the convolution mask used for the measurements 
and the local statistical properties of the noise (Bonnet \& 
Mellier 1995, Kaiser et al. 1995).  
Several critical issues of the method, related to the 
identification and delimitation of individual objects have
led Van Waerbeke et al. (1996)  to consider faint sources down
to the noise level as a global density field and to 
measure  the weak lensing effects from the analysis of the 
autocorrelation function of pixels (ACF) in CCD images. 
The concept is  simple and can be easily understood: the Fourier Transform
of an elliptical distribution of light is a conjugate elliptical
distribution of density in the $u,v$ plane  with the same ellipticity
(rotated by $\pi$/2). Thus the ACF on the angular scale of faint
distant galaxies is a new mathematical object that sum up all the 
$u,v$ sources  down to the noise  and
 immediately reveals the shear of all the sources in  the scanning
aperture.
An important point is that ACF avoids measurement of centroids 
and  shape parameters of individual galaxies which 
considerably reduces the uncertainties coming from
  sources of errors on the geometry of small and noisy objects.

\begin{figure}
\vskip 0cm
\hskip 7truecm
\psfig{figure=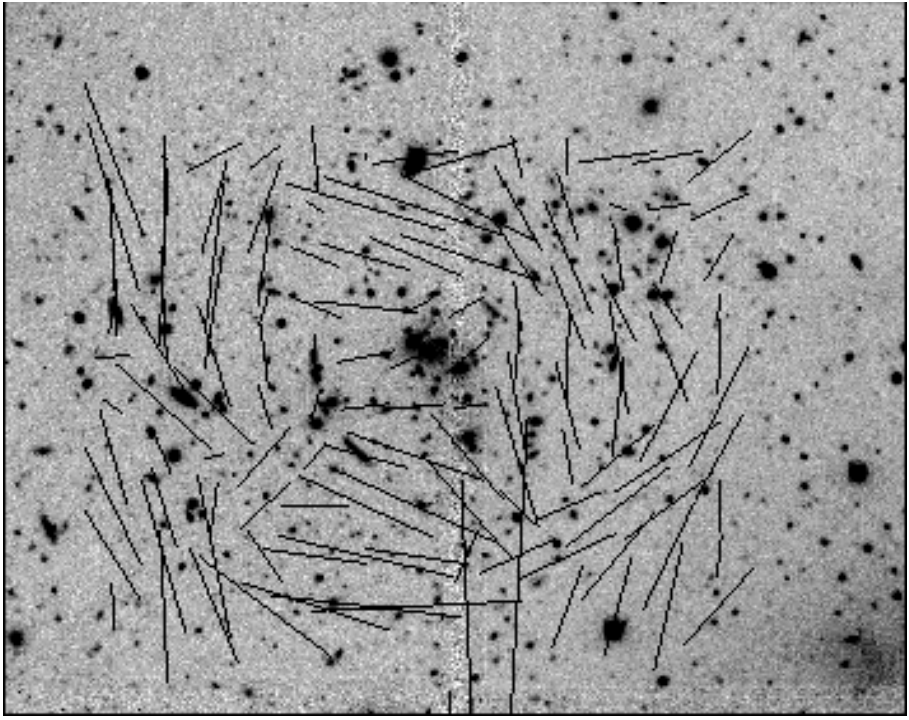,width=12. cm}
\caption{Shear map around Cl0024 from the ACF method obtained from a 
mosaic of two CCD images. The map can be
compared with the first one obtained by Bonnet et al. (1993). The
resolution is better and in particular the innermost pattern shows now that
the central mass distribution Cl0024 is bimodal.}
\end{figure}

Following Van Waerbeke et al. (1996) and Van Waerbeke \& Mellier (1996)
we can express the relation between the surface brightness, 
$I(\vec {\bf \theta})$, in the image plane at the position   
$\vec \theta$  and  the surface brightness in the source plane
$I^{(s)}$ by  
\begin{equation}
I(\vec \theta)=I^{(s)}({\cal A}\vec \theta),
\end{equation}
which can be extended
to the ACF (e.g. the
two-point autocorrelation function of the light distribution in a given
area),
\begin{equation}
\xi(\vec \theta)=\xi^{(s)}({\cal A}\vec \theta)\ ,
\end{equation}
 and the thin lens equation can be re-written  
as
\begin{equation}
\xi(\vec \theta)=\xi^{(s)}(\theta)-\theta \ \partial_{\theta}
\xi^{(s)}(\theta)
[1-{\cal A}]   \ .
\end{equation}
It proves that the local ACF behaves like a new object. In the image plane
 $\xi(\vec \theta)$  can be understood  as
the sum of an isotropic unlensed term,
$\xi^{(s)}(\theta)$, an isotropic lens
term which depends on $\kappa$, and an anisotropic term which depends on
$\gamma_i$.

The weak lensing information  is now given by
the shape matrix $\cal M$ of the ACF,
\begin{equation}
{\cal M}_{ij}={\int {\rm d}^2\theta\ \xi (\vec \theta)\ \theta_i\ 
\theta_j\over \int {\rm d}^2\theta\ \xi (\vec \theta)}\ .
\end{equation}
and the shape matrix in the image plane is simply related to the shape
matrix in the source plane ${\cal M}^{(s)}$ by 
${\cal M}_{ij}={\cal A}_{ik}^{-1} {\cal A}_{jl}^{-1} {\cal M}^{(s)}_{kl}$. 
If the galaxies
are isotropically distributed in the source plane,
$\xi^{(s)}$ is isotropic, and  ${\cal M}^{(s)}_{ij}=M\delta_{ij}$, 
where $\delta_{ij}$ is the identity matrix.
Using the expression of the amplification matrix $\cal A$ we can 
formally write  $\cal M$ as follows:
\begin{equation}
{\cal M}={M(a+|g|^2)\over (1-\kappa)^2(1-|g|^2)} 
\pmatrix{ 1+\delta_1 & \delta_2 \cr \delta_2 & 1-\delta_1 \cr }\ .
\end{equation}
Finally, the observable quantities (distortion $\delta_i$ and magnification
$\mu$) are given in terms of the components of the ACF shape matrix,
\begin{equation}
\delta_1={{\cal M}_{11}-{\cal M}_{22}\over {\rm tr}({\cal M})} \ ; \ \
 \
\delta_2={2{\cal M}_{12}\over {\rm tr}({\cal M})} \ ; \ \ \
\mu=\sqrt{{\rm det}({\cal M})\over M},
\end{equation}
where ${\rm tr}({\cal M})$ is the trace of $\cal M$ and
${\rm det}({\cal M})$ is the determinant of $\cal M$.

As for classical methods, we see that the
distortion is available from a direct measurement in the image plane
while
the magnification measurement requires to know the value of $M$ which
is related to the light distribution in the source plane or in any   
unlensed reference plane. 
The  main advantage of the ACF method is that it provides a new 
way to measure
$\delta_i$ and $\mu$ which does not depend upon the geometry
of the scanning aperture on the CCD image. Furthermore, 
 its signal to noise ratio is  
proportional to the number density $n$ of background galaxies:
$S/N \ \propto \ n $ instead of $S/N  \ \propto  \sqrt{n}$
for the standard methods (see figure 1).

A full description of the practical implementation of the
ACF and first results are given
in Van Waerbeke et al. (1996), Van Waerbeke \& Mellier (1996) and 
Mellier et al. (1996). Clearly, the ACF is potentially the most
powerful and promising 
technique to map very weak shears as those already detected 
at the outer periphery of clusters, those around bright 
quasars that seems  magnifed by (unseen) condensations of mass 
(Fort et al. 1996b) or those predicted by  large scale structures. 

As the signal to noise of the ACF increases as the galaxy number density,
the method will be very attractive on deep VLT images    
with good image qualities (seeing $<0.7"$). FORS should be
a unique instrument for some of these programmes.   
 
\section{The distance of background sources with $B>25$}
\subsection{Present status of the arc(et)s redshift distribution} 
Spectroscopic redshifts of luminous giant arcs allow
to calculate the angular distances $D_{\rm d}, 
D_{\rm ds}$ and $D_{\rm s}$ (see Eq. (1) of Schneider in this
proceedings) and  to get the absolute scaling of the lens potential. 
If arclets are also observed in regions where the potential are 
properly probed by the modelling of giant arcs it is then possible to infer
the most probable redshift of each individual arclet from
their position and their ellipticity.
The method has been successfully developped by
Kneib et al. (1994) using some cluster lenses as {\sl very low resolution
gravitational spectrograph}. The most probable redshift for each 
arclet is obtained by assuming it corresponds to the source plane where 
the distortion is minimum. Potentially, it allows to provide redshift 
($\delta  z \pm 0.1$) 
of equivalent un-magnified galaxies down to $B=27$ if the lens modelling 
is reliable and the shape
of each arclet is measurable with a good accuracy. 
Recently, Kneib et al. (1996) used this technique on deep HST images 
of A2218 and show that, as for A370 (Kneib et al. 1994), the colour-redshift 
diagram seems consistent with the deepest
redshift surveys. However, because it is a probabilistic approach, 
its efficiency must be also checked from spectroscopy.  This long term
programme started one year ago and spectroscopic surveys of the "brightest"
arclets in A2218, A2390 and others are underway. 
Ebbels et al. (1996) and B\'ezecourt \& Soucail (1996) purposely selected
arclets showing bright spots of stars
forming regions on HST images in order to detect 
an emission line and get secure redshift (see figure 2).
The rather good agreement between spectroscopic redshifts and lensing 
inversions demonstrate that gravitational redshift are reliable.
However, the observations
require a large amount of nights on 4 meter telescopes and cannot be
extended to the faintest arclets.

\begin{figure}
\vskip 0cm
\hskip 3truecm
\psfig{figure=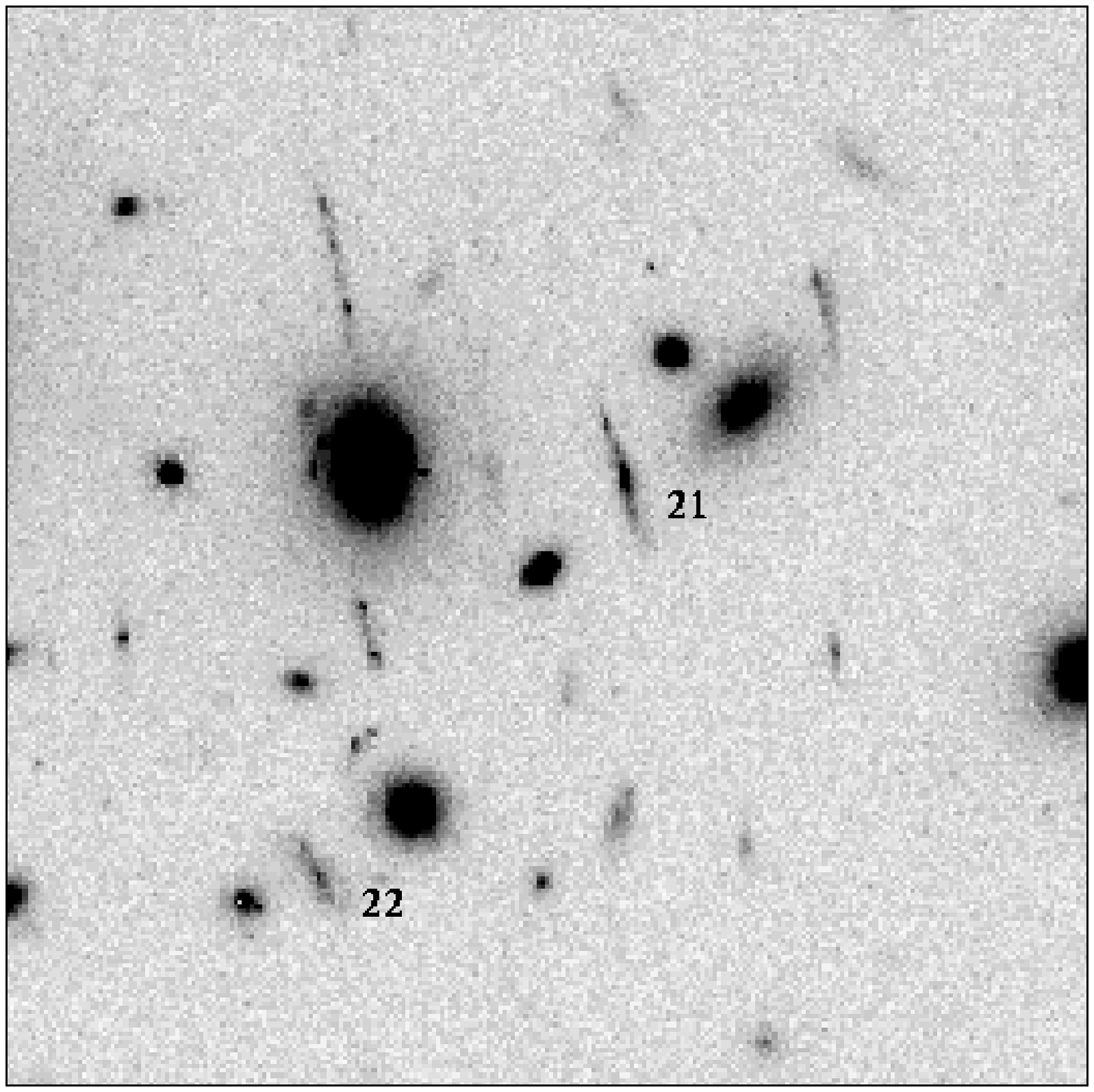,width=10. cm}
\psfig{figure=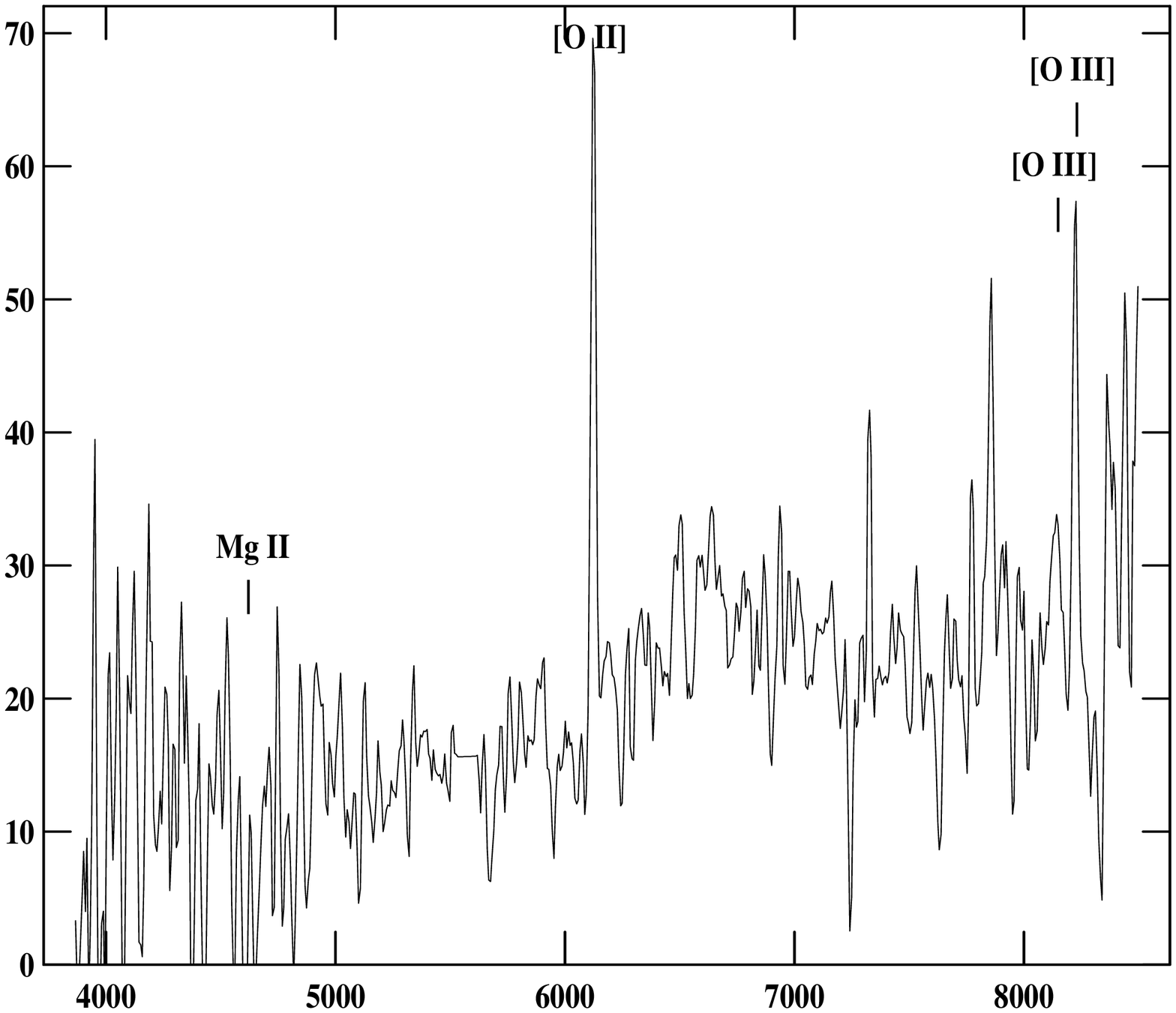,width=6. cm}
\psfig{figure=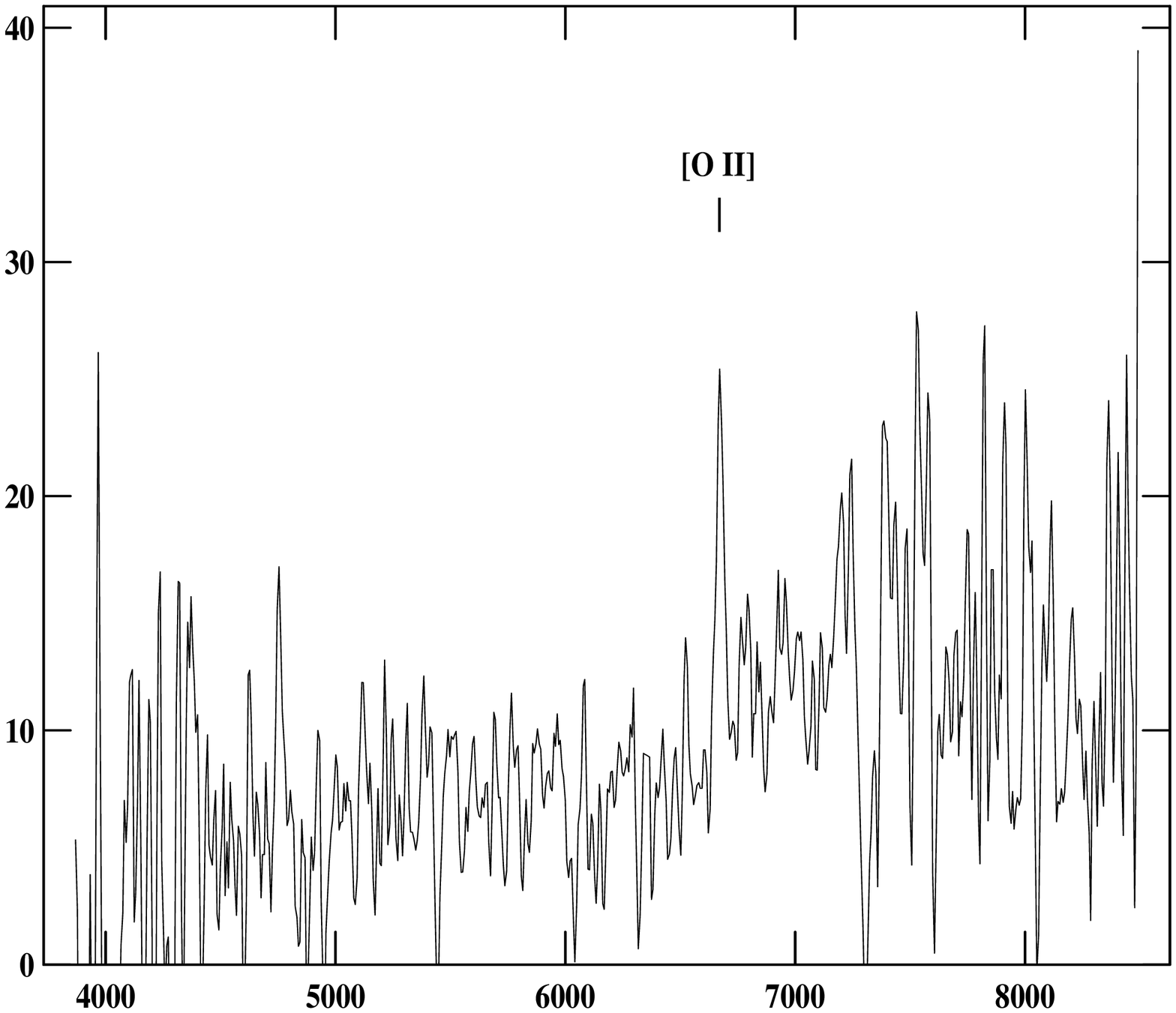,width=6. cm}
\caption{Spectroscopy of faint arclets in A2390. The top panel is a deep
HST image of its central region. The arclets are clearly visible with 
some of them showing image multiplicity with image parity changes. The arclets
21 and 22 have been observed by B\'ezecourt and Soucail (1996) with the
MOS multiobject spectrograph at CFHT. As expected, they show the  
[OII]$\lambda$3727 emission line from which the redshifts are easily 
measured   ($z=0.643$ and $z=0.790$ respectively. Courtesy J. B\'ezecourt).}
\end{figure}
The spectroscopic observation of arc(et)s for the strong lensing modelling,
 for future cosmological tests, as well as for
the study of the distribution and evolution of faint distant 
galaxies will be continued with the VLT because one can reasonnably go at 
least 0.5 magnitude beyond 4 meter telescopes. In fact, one may be able to
go a bit deeper by 
using FORS with the Va-et-Vient spectroscopic mode proposed by 
Cuillandre et al. (1994). This technique works well in low resolution
spectroscopy in particular to derive the  spectral energy   
distribution  of faint distant galaxies without
emission lines  or QSOs with broad emision lines (for galaxies with thin 
emission line 
higher spectral resolution may be more suited). According to the gain 
estimated by Cuillandre et al., the flat field residuals are removed with
a much better accuracy than in standard  spectroscopic observations and
it is possible to go 1 magnitude deeper.  
In that case, FORS with
the Va-et-Vient spectroscopic mode could observe arclets as faint as 26.5
which corresponds to 20-30 arclets per cluster in three hours.

Since the lensing magnification is large, redshift surveys of arc(let)s 
could probe the redshift distribution of galaxies with $B>24$. But
unfortunately,   the galaxy sample is biased in three ways. First, only arclets 
with star forming emisson lines are
selected. Second, sky features and redshift effect conspire to offer
peculiar windows of redshift visibility. Finally, as it is discussed 
in the next section, the
magnification bias also favours observations of blue galaxies rather than red.
So, at present the spectroscopy of arclets is  
crucial for the lens modelling but the redshift
distributions obtained from these methods are still questionnable.

\subsection{Probing the redshift of sources up to $B=28$ with the
magnification bias}

Broadhurst (1995) first demonstrated that the radial surface density 
of background galaxies around a cluster lens like
Abell 1689 varies according to the predicted magnification bias of the lens and
can be used to measure $\kappa$ directly and to map
the projected mass density of the cluster.
More attractive, this (anti) magnification bias effect provides a good 
way to break the intrinsic degeneracy
of the inversion methods based on gravitational weak shears:  
 an additionnal plan of constant mass density
on the line of sight does not change the shear pattern but just
the convergence of light beams.

The radial behavior of the so-called Broadhurst's effect results from
the competition between the gravitational magnification that increases 
the detection of individual objects above the limit of detection and the  
deviation
of light beam that decreases the apparent number counts. 
Therefore the amplitude of the magnification bias 
depends explicitly on the slope of 
the galaxy counts as a function of magnitude and
on the magnification factor of the lens:
\begin{equation}
   N(<m,r) = N_0(<m) \ \mu(r)^{2.5\gamma-1} \ ,
\end{equation}
 where  $\mu(r)$ is the magnification factor of the lens, $N_0(<m)$ the
intrinsic number density in a nearby empty field and
  $\gamma$  is the intrinsic count slope:
\begin{equation}
\gamma = {dlogN(<m) \over dm }\ .
\end{equation}
The radial magnification bias $N(<m,r)$ shows up only when
the slope $\gamma$ is different
from the value 0.4; otherwise, the increasing number of
magnified sources is exactly cancelled by the apparent field dilatation
and there is no effect on $N(<m,r)$.
As noticed by Broadhurst, a radial amplification bias cannot
be observed 
for $B(<26)$ since
the slope is almost this critical  value (Tyson 1988, Shanks 1996) but it
can be detected in the R or I bands when the slopes are close to 0.3
(Smail et al. 1995).

\begin{figure}
\vskip 0cm
\hskip 3truecm
\psfig{figure=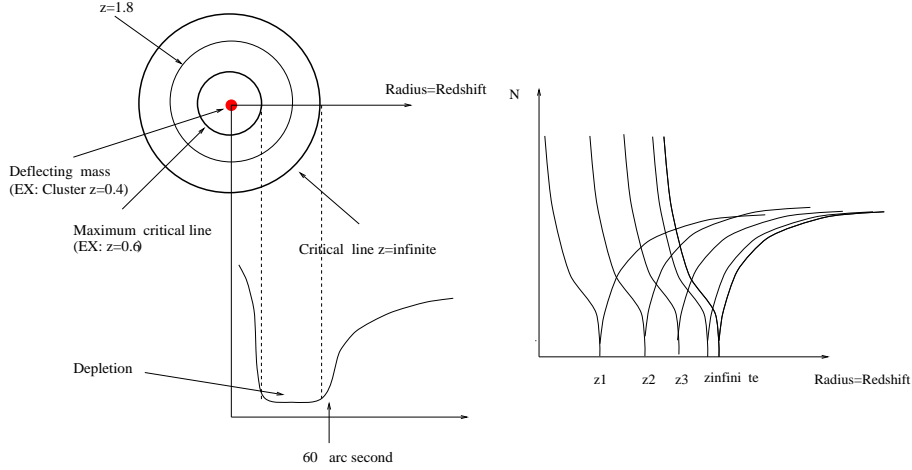,width=12. cm}
\caption{Measurement of redshifts from depletion curves. The
left panel shows the depletion by a singular isothermal sphere 
as it would be observed on the sky 
(top left) and the radial projected density of galaxies (bottom left). 
For a single source redshift, when the lens is perfectly known, 
the minimum of a depletion curve is sharp and its radial position is 
formally equivalent to a redshift (for example the position $z1$ of the
right panel). The radial position of the miminum increases with the
redshift of sources but the depletion curves tighten and converge
towards the curve corresponding to sources at infinity.  In a realistic case, 
the redshift distribution is broad and the individual curves must 
be added. The bottom left  panel shows the depletion as it
would be observed: instead of
the single peaked depletion we expect a more pronounced minimum  between
two radii (i.e. two redshifts ) whose angular positions depend on
the
cosmological constant for high redshift sources. Thus, if the mass
distribution of the lens is well known as in the rich lensing cluster
Cl0024+1654, the distribution of sources and 
$\lambda$ can be inferred from the shape of the depletion
curve.}
\end{figure}
When $\gamma < 0.3$ a sharp decrease of the number of galaxies
is expected in regions of strong magnification
 close to the critical radius of the lens corresponding to the redshift of the
background sources.
Since the critical radius increases with redshift, every population of
galaxies at different redshifts 
will display a distinct peaked annular depletion at its own critical line. 
 It can result a shallower depletion between the smallest and 
the largest critical line
which depends on the redshift  distribution of the galaxies.
(Figure 3 and 4). In short, the lens gives us a new way
to sort out different class of galaxies versus their redshift
distribution. It was first used by Fort et al (1996a) with the cluster 
Cl0024+1654 to study the
faint distant galaxies population in the extreme range of magnitude
$B=26.5-28$ and $I=25-26.5$ after a detection of the sources in the 
sky background noise. For this selected bins of magnitude they
found on their CFHT blank fields  
that the counts slope was near 0.2, well suited for the study of
the Broadhurst's effect. After analysis of the shape of the depletion 
curve (figure 4), $60\% \pm 10\%$ of the
$B$-selected galaxies were found
between $z=0.9$ and $z=1.1$ while most of the remaining $40\%$
galaxies appears to be broadly distributed around a redshift of   
$z=3$. The
$I$ selected population present a similar distribution  with two maxima, but 
spread up to a larger redshift range  with about 20\% above  $z > 4$.
In fact, many of the $I$ selected galaxies
were not detected in $B$ as if their Lyman $\alpha$ discontinuity
has already crossed the $B$ filter band 
(redshift range  $z>3.5$).  

The main characteristic of this observation is the long exposure time
with an excellent seeing condition. Otherwise the detection of such
faint objects with an accuracy almost comparable to the HDF would be
impossible. In that sense, it is quite clear that such
observations are perfectly suited for FORS during the best
seeing period on Paranal.  So far it seems that the magnification bias 
 is probably a good technique to study the abundance of galaxies at
very large distance. Just one limitation comes from the fact that
the critical lines of a gravitational lens tend to rapidly merge with 
the critical line at infinity for a redshift larger than 3. So we can just 
estimate the relative  population of faint galaxies above this redshift. 
Conversely, this limitation was in turn
used by Fort et al (1996a) for a first attempt of a direct  
measurement of the value of the cosmological constant.

\section{Measuring the cosmological constant from depletion curves}

\begin{figure}
\vskip 0cm
\hskip 3truecm
\psfig{figure=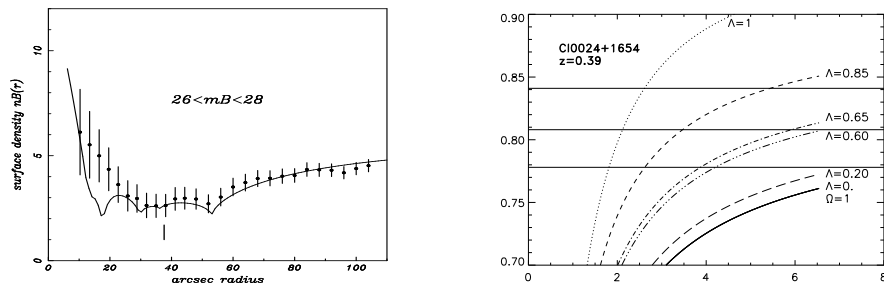,width=12. cm}
\caption{A measurement of the cosmological constant from a depletion 
curve.  The left panel shows the depletion curve measured in
Cl0024+1654. By using  the redshift corresponding to the two limits of 
the plateau observed on the curve,
 one can constrain the cosmological constant from the    
position of the second minimum. Whatever the
redshift of the most distant sources visible on the images, we see that
the angular position where the depletion curve raises again imposes that
$\lambda>0.65$}
\end{figure}
The qualitative profile of the depletion curve in a
direction perpendicular to the critical line is given on figure 3.
 When a distribution of detectable galaxies spread up to large 
redshifts all the most 
distant ones beyond $z=3$ will form their sharp depletion at 
almost the same
radial distance that, for simplicity,  we identify here
to $rc_{\infty}$. The most abundant the very distant galaxies  are,
the
stronger is the rising discontinuity at the end of the depletion
(figure 4). Basically, Fort et al (1996a) have  used the modelling of the
lens inferred from the giant arc and the location of the last critical line 
$R_I$  detected in the $I$ bandpass
to implement a cosmological test on the value of the cosmological constant. 
For a flat universe they
found  $rc_{\infty}$ at a too large distance from the center of 
the cluster for a non zero cosmological constant\footnote{except if there is
an additional, and so far unseen, deflector hidden behind Cl0024+1654.}. 
With a  thin lens hypothesis and despite 
 large uncertainties their observations seems to  
favour an $\Omega_{\Lambda}$-dominated flat universe with 
a cosmological constant ranging from 0.6 to 0.9. On the other 
hand, statistics of gravitational lens events on QSOs seems to 
rule out models with  $\Omega_{\Lambda}> 0.6$ (Kochanek 1995).  
 Thus, if the conclusions of the magnification bias and of 
the QSO statistics are correct, 
few room is left for a possible value of  the cosmological constant.  

 The Fort et al. (1996a)  preliminary result  is still questionable but
at least it demonstrates that the  magnification bias is worth to be explored
further with many clusters, both to study the distribution and evolution of 
faint distant galaxies and  to provide cosmological tests 
 in clusters of galaxies with large multiple arcs. A reasonnable sample of 20
rich lensing clusters with simple geometry observed with FORS on the VLT
would provide very deep exposures (to $B \approx 28-29$) rapidly with 
a very low flat field residuals which is crucial for detection in the
noise. It will be a challenging programme with FORS on the VLT.

\section{Conclusion}
It is now established from the weak lensing analyses we have described here 
 as well as  from the HDF data and recent deep spectroscopic surveys
(this conference) that it is possible to detect and study the
population of galaxies at redshift beyond  
$z=4$. Most of them, despite probably strong star forming activities,
will be close to or within the sky background noise. It is
already demonstrated with 4 meter class telescopes that their detectability is 
not a big challenge for VLTs during period of excellent seeing.
Their observations on a regular basis is
a tremendous chance for weak lensing studies 
and their large cohort of exciting applications.  However, 
for the measurement of very faint shear ($\approx 1\%$), 
 it is crucial to control the stability
of the image quality to a level which was never thought of before.
 A dedicated and a large comprehensive effort must be done in this domain 
by instrument builders and observers in order  to reach the 
ultimate imaging capabilities of the VLTs.

\section*{Acknowledgments}
We thank J.-C. Cuillandre, R. Pell\'o, P. Schneider, C. and S. Seitz,
and  L. Van Waerbeke for  stimulating discussions about lensing and 
prospective aspects. J. B\'ezecourt kindly provided the spectra shown 
in this review.
%
%


\begin{thebibliography}
%
%
\bibitem{}{}{}
B\'ezecourt, J., Soucail, G. 1996, SISSA preprint astro-ph/96006064.
\bibitem{}{}{}
Bonnet, H., Mellier, Y., Fort, B. 1994, ApJ 427, L83.
\bibitem{}{}{}
Bonnet, H., Mellier, Y. 1995, A\&A 303, 331.
\bibitem{}{}{}
Broadhurst, T., Taylor, A.N., Peacock, J. 1995, ApJ 438, 49.
\bibitem{}{}{}
Broadhurst, T. 1995, SISSA preprint astro-ph/9511150.
\bibitem{}{}{}
Cuillandre, J.-C., Fort, B., Picat, J.-P., Soucail, G., Altieri, B.,
Beigbeder, F., Dupin, J.-P., Pourthi\'e, T., Ratier, G. 1994, A\&A 281,
603.
\bibitem{}{}{}
Ebbels, T. M. D., Le Borgne, J.-F., Pell\'o, R., Ellis, R. S., Kneib,
J.-P., Smail, I., Sanahuja, B. 1996, SISSA astro-ph/9606015.
\bibitem{}{}{}
Fort, B., Mellier, Y. 1994, A\&A Review 5, 239, 292.
\bibitem{}{}{}
Fort, B., Mellier, Y., Dantel-Fort, M. 1996a, SISSA preprint
astro-ph/9606039.
\bibitem{}{}{}
Fort, B., Mellier, Y., Dantel-Fort, M., Bonnet, H., Kneib, J.-P. 1996b,
A\&A 310, 705.
\bibitem{}{}{}
Kneib, J.-P., Mathez, G., Fort, B., Mellier, Y., Soucail, G.,
Longaretti, P.-Y. 1994, A\&A 286, 701.
\bibitem{}{}{}
Kneib, J.-P., Ellis, R. S., Smail, I., Couch, W. J., Sharples, R. M.
1996, SISSA preprint astro-ph/9511015.
\bibitem{}{}{}
Kochanek, C. S. 1995, SISSA preprint astro-ph/9510077.
\bibitem{}{}{}
Mellier, Y., Van Waerbeke, L., Bernardeau, F., Fort, B. 1996, Proceedings of 
the  VIII$th$ Rencontres de Blois "{\sl Neutrinos, Dark Matter and the 
Universe}". Blois, France 1996. 
\bibitem{}{}{}
Narayan, R., Bartelmann, M.  1996, SISSA preprint astro-ph/9606001
\bibitem{}{}{}
Shanks, T. 1996, Proceedings of the 37$^{th}$ Herstmonceux
 Conference "{\sl HST and the High Redshift Universe}". Cambridge 1996.
N. Tanvir, A. Arag\'on-Salamanca, J. V. Wall eds.
\bibitem{}{}{}
Smail, I., Hogg, D., Yan, L., Cohen, J. G. 1995, ApJ 449, L105.
\bibitem{}{}{}
Tyson, A. J. 1988, AJ 96, 1.
\bibitem{}{}{}
Tyson, J. A., Valdes, F., Wenk, R. 1990, ApJ 349, L1.
\bibitem{}{}{}
Van Waerbeke, L., Mellier, Y., Schneider, P., Fort, B., Mathez, G.
1996, A\&A in press. SISSA preprint astro-ph/9604137.
\bibitem{}{}{}
Van Waerbeke, L., Mellier, Y. 1996, Proceedings of the XXXIst
Rencontres de Moriond "{\sl Dark Matter in Cosmology. Quantum
Measurement, Experimental Gravitation}". Les Arcs, France 1996. SISSA preprint
astro-ph/9606100.

%
\end{thebibliography}
\end{document}